\documentclass[9pt, sigconf,bookmarks=false,screen]{acmart}
\AtBeginDocument{%
  }
    
\usepackage[utf8]{inputenc} %
\usepackage[T1]{fontenc}    %

\usepackage{pifont}
\usepackage{xcolor}         %
\usepackage{booktabs}
\usepackage{hyperref}
\hypersetup{colorlinks,
    linkcolor=red,citecolor=citecolor,urlcolor=blue}
\usepackage{algorithm}
\usepackage{graphicx}
\usepackage{threeparttable}
\usepackage{multirow}
\usepackage{amsmath}
\usepackage{bm}
\usepackage{bbm}
\usepackage{amsfonts}
\usepackage{amsthm}
\usepackage[]{algpseudocode}
\usepackage{enumitem} %
\usepackage[subrefformat=parens,labelformat=parens]{subfig}
\usepackage{colortbl}

\usepackage{wrapfig}

\setcitestyle{sort&compress,numbers}
\usepackage{xspace}

\definecolor{citecolor}{RGB}{34,139,34}
\definecolor{mydarkblue}{rgb}{0,0.08,1}
\definecolor{mydarkgreen}{rgb}{0.02,0.6,0.02}
\definecolor{mydarkred}{rgb}{0.8,0.02,0.02}
\definecolor{mydarkorange}{rgb}{0.40,0.2,0.02}
\definecolor{mypurple}{RGB}{111,0,255}
\definecolor{myred}{rgb}{1.0,0.0,0.0}
\definecolor{mygold}{rgb}{0.75,0.6,0.12}
\definecolor{myblue}{rgb}{0,0.2,0.8}
\definecolor{mydarkgray}{rgb}{0.,0.2,0.2}

\definecolor{lightred}{RGB}{255,235,235}
\definecolor{lightgreen}{RGB}{235,255,235}
\definecolor{lightblue}{RGB}{235,235,255}
\definecolor{lightcyan}{RGB}{235,255,255}
\definecolor{lightmagenta}{RGB}{255,235,255}
\definecolor{lightyellow}{RGB}{255,255,235}

\definecolor{qxkcolor}{RGB}{215,235,255}
\definecolor{softmaxcolor}{RGB}{230,235,255}
\definecolor{probxvcolor}{RGB}{255,255,235}

\definecolor{topkcolor}{RGB}{255,235,235}
\definecolor{zecolor}{RGB}{255,255,235}
\definecolor{dynacolor}{RGB}{235,255,255}

\definecolor{reviewcolor}{RGB}{0,0,200}

\renewcommand{\vec}[1]{\boldsymbol{#1}}

\theoremstyle{plain}

\theoremstyle{definition}

\algdef{SE}[DOWHILE]{Do}{doWhile}{\algorithmicdo}[1]{\algorithmicwhile\ #1}%

\newcommand{\squishlist}{
 \begin{list}{$\bullet$}
  { \setlength{\itemsep}{0pt}
     \setlength{\parsep}{3pt}
     \setlength{\topsep}{3pt}
     \setlength{\partopsep}{0pt}
     \setlength{\leftmargin}{1.5em}
     \setlength{\labelwidth}{1em}
     \setlength{\labelsep}{0.5em} } }

\newcommand{\squishend}{
  \end{list}  }

\newcommand{\name}{\texttt{Apollo}\xspace}

\graphicspath{{./figs/}}
\usepackage{footnote}

\usepackage{geometry}
\usepackage{tabularx,booktabs}
\newcolumntype{L}[1]{>{\raggedright\arraybackslash}p{#1}}
\newcolumntype{C}[1]{>{\centering\arraybackslash}p{#1}}
\newcolumntype{R}[1]{>{\raggedleft\arraybackslash}p{#1}}
\newcolumntype{Y}{>{\centering\arraybackslash}X}

\copyrightyear{2025} 
\acmYear{2025} 
\setcopyright{acmcopyright}\acmConference[ICCAD '25]{IEEE/ACM International Conference on Computer-Aided Design}{October 26--30, 2020}{Munich, Germany}
\acmBooktitle{IEEE/ACM International Conference on Computer-Aided Design (ICCAD '25), October 26--30, 2025, Munich, Germany}
\acmPrice{15.00}
\acmISBN{979-8-4007-1293-7/25/03}

\begin{document}

\expandafter\def\expandafter\normalsize\expandafter{%
    \normalsize%
    \setlength\abovedisplayskip{0pt}%
    \setlength\belowdisplayskip{8pt}%
    \setlength\abovedisplayshortskip{-8pt}%
    \setlength\belowdisplayshortskip{2pt}%
}

\settopmatter{printacmref=false} %
\renewcommand\footnotetextcopyrightpermission[1]{} %
\pagestyle{plain} %

\newcommand*{\affaddr}[1]{#1} %
\newcommand*{\affmark}[1][*]{\textsuperscript{#1}}

\title{
Apollo: \underline{A}utomated R\underline{o}uting-Informed \underline{P}lacement for \underline{L}arge-Scale Ph\underline{o}tonic Integrated Circuits
}

\author{
Hongjian Zhou\textsuperscript{1},
Haoyu Yang\textsuperscript{3},
Nicholas Gangi\textsuperscript{2},
Zhaoran (Rena) Huang\textsuperscript{2},
Haoxing Ren\textsuperscript{3},
Jiaqi Gu\textsuperscript{1,$\dagger$}\\
\textsuperscript{1}Arizona State University \quad
\textsuperscript{2}Rensselaer Polytechnic Institute \quad
\textsuperscript{3}NVIDIA Corporation \\
\textsuperscript{$\dagger$}jiaqigu@asu.edu
}

\begin{abstract}
\label{abstract}
As technology advances, photonic integrated circuits (PICs) are rapidly scaling in size and complexity, with modern designs integrating thousands of components to meet the demands of artificial intelligence (AI), high-performance computing, and chip-to-chip optical interconnects. 
However, the analog custom layout nature of photonics, the curvy waveguide structures, and single-layer routing resources impose stringent physical constraints, such as minimum bend radii and waveguide crossing penalties, which make manual layout the \emph{de facto} standard.
This manual process takes weeks to complete and is error-prone, which is fundamentally unscalable for large-scale PIC systems.
Existing automation solutions have adopted force-directed placement on small benchmarks with tens of components, with limited routability and scalability.
To fill this fundamental gap in the electronic-photonic design automation (EPDA) toolchain, we present \name, the first GPU-accelerated, routing-informed placement framework tailored for large-scale PICs.
\name features an asymmetric bending-aware wirelength function with explicit modeling of waveguide routing congestion and crossings to preserve enough routing spacing for routability maximization.
Meanwhile, conditional projection is employed to gradually enforce a variety of user-defined layout constraints, including alignment, spacing, etc. 
This constrained optimization is accelerated and stabilized by a custom blockwise adaptive Nesterov-accelerated optimizer, ensuring stable and high-quality convergence. 
To catalyze research in PIC layout automation, we also develop and open-source large-scale PIC benchmarks derived from real-world photonic tensor core designs.
Compared to existing methods, \name can generate high-quality layouts for large-scale PICs with an average routing success rate of 94.79\% across all
benchmarks within minutes. 
By tightly coupling placement with physical-aware routing, \name establishes a new paradigm for automated PIC design—bringing intelligent, scalable layout synthesis to the forefront of next-generation EPDA.
Our code is open-sourced at \href{https://github.com/ScopeX-ASU/Apollo}{link}\footnote{\href{https://github.com/ScopeX-ASU/Apollo}{https://github.com/ScopeX-ASU/Apollo}}.

\end{abstract}

\maketitle
\vspace{-5pt}
\section{Introduction}
\label{sec:Introduction}

Integrated photonics has emerged as a transformative platform for high-performance computing~\cite{zhou2022photonic, NP_NATURE2017_Shen, NP_Nature2025_Hua, NP_Nature2025_Ahmed} and communication systems~\cite{NP_CICC2024_Wang,NP_DATE2020_Thonnart}. 
By manipulating and processing light signals on-chip through optical components, photonic integrated circuits (PICs) offer unparalleled advantages in speed, parallelism, and energy efficiency. 
These properties make PICs particularly attractive for a wide range of applications, from chip-to-chip optical interconnects~\cite{NP_CICC2024_Wang,NP_DATE2020_Thonnart,NP_TCAD2013_Ye,NP_ICCAD2022_Taheri,10323627} and data center networking, to artificial intelligence (AI) acceleration~\cite{NP_NATURE2017_Shen, NP_Nature2025_Hua, NP_Nature2025_Ahmed}, quantum computing, and optical signal processing.

Despite their immense potential, the physical layout of PICs remains largely a manual and iterative process, even in state-of-the-art commercial and academic design flows~\cite{korthorst2023photonic}.
Unlike digital VLSI, where placement and routing are decoupled and highly automated, PIC layout presents fundamentally different challenges due to the analog nature of light propagation and the unique physical constraints needed to maintain signal integrity, performance, and fabrication yield.
In PICs, we claim that waveguide routing must be tightly integrated into the placement process, far more so than in digital VLSI, because the feasibility and quality of routing are highly sensitive to component placement.
At the core of these challenges lies the fact that photonic waveguides are directional, curvilinear, and highly space-consuming~\cite{PD_ISPD2025_Zhou}. 
Unlike metal wires in VLSI circuits or on PCBs, waveguides must conform to minimum bend radii to avoid excessive propagation loss and cannot intersect with photonic devices. 
Waveguide crossings, analogous to vias in VLSI, must also be carefully planned to minimize crosstalk, insertion loss, and excessive detours, all of which are tightly influenced by placement. Crossings further introduce non-negligible area overhead and impose strict port orientation requirements, complicating the layout even more.
Moreover, PICs are typically restricted to a single layer (or a few for advanced PICs), resulting in severe layout resource contention.
These properties make waveguide routing a dominant constraint during placement. 
Furthermore, PICs often involve mixed-size components, ranging from compact 2$\times$2 $\mu m^2$ devices (e.g., micro-disk modulators and y-branches) to large 100$\times$1000 $\mu m^2$ devices (e.g., Mach-Zehnder modulators, MZM). 
Some devices exhibit extreme aspect ratios (e.g., 1:20), posing additional challenges for placement that must handle geometric diversity, routability, and photonic-specific layout rules.

Due to these complex requirements, current commercial PIC layout tools often rely on schematic-driven layout practices. Designers manually translate schematics into physical layouts, carefully placing components and routing waveguides to ensure correct port alignment, sufficient spacing between components to reduce crosstalk, and reserving enough room to accommodate bends and crossings. While this manual methodology can yield valid designs, it is highly time-consuming and cannot scale to large systems with thousands of photonic components.

Prior works have attempted to address PIC placement through force-directed approaches ~\cite{von2016platon, chen2025cponoc}. These methods are fully automated and generate placements without leveraging human design expertise or schematic guidance, resulting in layouts that are unintuitive and difficult for engineers to interpret or validate. While these works primarily focus on minimizing the number of waveguide crossings, they fail to account for critical routability considerations such as bend radii, port accessibility, and the area overhead of crossings. As a result, their placement solutions frequently lead to routing failures and do not yield physically legal layouts.

In this work, we propose \name, an automated, GPU-accelerated, routing-informed placement engine for large-scale PICs. 
Our framework directly tackles the inherent limitations of prior solutions by integrating physical routing constraints and designer-specified layout rules into the placement phase. \name models waveguide routing as a first-class entity, enabling layout-aware placement that explicitly considers bend radii, crossing penalties, and port orientations. 
In contrast to prior methods that neglect practical layout constraints, \name ensures high-quality layouts that are physically routable.

\begin{itemize}
  \item \textbf{Routing-Informed Placement Engine}: Co-optimizes bending-aware wirelength and routing congestion-driven component distribution using a unified differentiable framework.
  \item \textbf{Bending-Aware Wirelength and Spacing Estimation}: Accurately models waveguide geometries and local congestion to preserve layout feasibility.
  \item \textbf{Progressive Constraint Handling}: Handles spacing, alignment, symmetry, and uniformity using a conditional projected gradient descent strategy.
  \item \textbf{Blockwise Adaptive Nesterov Optimizer}: Ensures stable and high-quality convergence on mixed-size PIC designs via parameter group-wise gradient updates.
  \item \textbf{Superior Routability}: Compared to existing methods, our \name can generate high-quality layouts for large-scale PICs with an average routing success rate of 94.79\% across all benchmarks within minutes. 

\end{itemize}

\vspace{-5pt}
\section{Preliminaries}
\label{sec:Preliminaries}
In this section, we review the background and the motivation.
\subsection{Photonic Circuit Placement}

Commercial PIC physical design toolflow has traditionally relied on schematic‑driven layout methodology~\cite{korthorst2023photonic}, in which devices are placed according to their logical positions and then manually abutted via waveguide ports, a time‑consuming process that does not scale to large systems. 
To boost productivity, the work~\cite{hendry2011vandal} introduced a semi‑automatic Photonic CAD tool for visual place‑and‑route of on‑chip photonic networks. 
Fully automatic frameworks such as Proton~\cite{proton} and Platon~\cite{von2016platon} then focused on minimizing insertion loss, particularly at crossings, while another work~\cite{chuang2018planaronoc} further reduced loss through device flipping and rotation. 
A recent research incorporated critical‑path insertion‑loss optimization directly into placement~\cite{chen2025cponoc}.
Another work~\cite{truppel2019psion} jointly optimized network topology and its physical realization. Thermal‑aware approaches have also been proposed~\cite{jiao2018thermal}. However, these efforts largely overlook the geometric cost of waveguide routing itself. 
In this work, we argue that a routing‑informed placement engine, one that explicitly accounts for bends, crossings, and congestion during device placement, is essential for truly scalable, routable PIC layout.
 
\vspace{-5pt}
\subsection{Analytical Mixed-Size Placement}
Analytical placement usually consists of three stages: global placement (GP), legalization (LG), and detailed placement (DP). Global placement spreads out instances in the layout; legalization removes the remaining overlaps between instances and aligns instances to placement sites; detailed placement performs incremental refinement to further improve the quality. Since the quality of the final placement solution largely depends on the global placement and legalization stage, we mainly focus on them in this work.

Global placement aims at minimizing the wirelength cost subject to density constraints. The density constraints are relaxed into a density penalty term, computed as the potential energy of an electrostatic system where cells are modeled as charges, as in ePlace~\cite{PLACE_TODAES2015_Lu}. The problem can be formulated as:
\begin{equation}
\label{eq:Placement}
    \min_{(x, y)} \sum_{e \in \mathcal{E}} w_e \cdot WL(e; x, y) + \lambda \cdot D(x, y),
\end{equation}
where $\mathcal{E}$ is the set of nets, $(x, y)$ are the coordinates of all the instances, $w_e$ is the weight of net $e$, $WL(\cdot)$ is a differentiable wirelength cost function, and $D(\cdot)$ denotes the density penalty that spreads instances out in the layout. The non-overlapping constraint can be satisfied by gradually increasing the density weight $\lambda$, e.g., using the Lagrangian method.
From an optimization perspective, this formulation can be extended to mixed-size placement as well, converting the analytical mixed-size global placement problem into an unconstrained optimization problem with a differentiable objective function.
Our \name will use this analytical framework to co-optimize waveguide routing lengths and congestion while ensuring device non-overlapping and other user-defined layout constraints in a scalable way.

\vspace{-5pt}
\section{Apollo: Automated PIC Placement Framework}
\label{sec:Method}

In this section, we present the details of our PIC placer \name.

\vspace{-5pt}
\subsection{Understanding the Placement Challenges of Photonic Circuits}
\label{sec:Challenge}
Photonic IC layout presents fundamentally different challenges from digital VLSI due to the analog, custom-layout nature of photonics and stringent physical constraints. 
In PICs, placement and routing are tightly coupled.
Layout feasibility and quality are highly sensitive to component positions, port orientations, and available spacing.
This tight coupling arises from several critical factors:

\ding{202} \emph{Photonic waveguides are directional, curvy, and space-consuming.} Unlike flexible metal wires in VLSI, optical waveguides are of large width (500-1000 nm), require large minimum bend radii (5-10 $\mu m$) to avoid optical loss, and cannot intersect with other devices (unlike PCB routing), which are themselves optical waveguide structures. 
As a result, routing becomes a dominant constraint during placement.

\ding{203} \emph{PICs are limited to very few routing layers.} This leads to scarce routing resources and makes 90-degree waveguide crossings topologically inevitable in many designs, introducing insertion loss, crosstalk, and significant area overhead (as crossings take spaces like via in VLSI/PCB), further increasing congestion and routability complexity.

\begin{figure}
    \centering
    \includegraphics[width=\columnwidth]{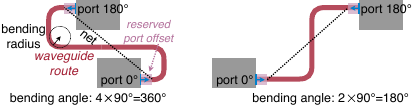}
    \vspace{-10pt}
    \caption{Solutions with the same HPWL give different waveguide routing and total bending angles.}
    \label{fig:angle}
     \vspace{-10pt}
\end{figure}

\ding{204} \emph{Ports are orientation-sensitive and often densely packed.} Improper port orientation can introduce excessive bends and detours as shown in Fig.~\ref{fig:angle}, which frequently result in unroutable or lossy connections, shrinking the feasible routing solution space and necessitating orientation-aware placement strategies.
Moreover, photonic devices often feature high port density; for instance, a multimode interference (MMI) device or star coupler may include 16 input and 16 output ports tightly packed along its edge. 
These dense configurations implicitly require substantial surrounding space to accommodate waveguide escape routes and crossings, further complicating placement.

\ding{205} \emph{Mixed-size PIC components are geometrically diverse.} Photonic components span from compact 2$\times$2 $\mu m^2$ compact splitters/resonators to millimeter-scale high-speed modulators with extreme aspect ratios (e.g., 1:20). 
This diversity mirrors the challenges of mixed-size VLSI placement, long known to be combinatorial and ill-conditioned for joint optimization.

\ding{206} \emph{Lack of high-quality PIC routers.} In PIC layout, placement quality is tightly coupled with the capabilities of the router. 
However, the current EPDA ecosystem lacks mature, high-quality routers for PICs. This practical limitation means that many routing complexities~\cite{PD_ISPD2025_Zhou, 10.1145/3725888}, such as bend management, crossing avoidance, and port alignment, must be preemptively handled during placement. 
As a result, the burden of achieving a routable and high-performance layout is currently disproportionately shifted to the placement engine.

These challenges collectively demand a routing-informed placement framework that models photonic-specific constraints from the outset.
Our \name framework addresses these challenges by incorporating routing-aware component spacing and bending-aware wirelength modeling, ensuring high-quality and physically routable layouts for large-scale PICs.

\subsection{Overview of \name Framework}
\label{sec:Overview}
\name consists of an analytical global placement engine with a simple legalization stage.
It is built on an open-source GPU-accelerated VLSI placement framework, DREAMPlace, that enjoys fast PyTorch-based programming and automatic differentiation engines for gradient-based optimization.
It features a LEF/DEF-inspired PIC benchmark suite in YAML format based on real large-scale photonic tensor core designs, which is intentionally designed to support seamless integration with open-source PIC layout tools \texttt{GDSFactory} and an open-source automated PIC router \texttt{LiDAR} for end-to-end evaluation.
Different from the standard LEF/DEF definition, we extend it to incorporate the port orientation of photonic components with targeted waveguide width and crossing sizes.
Different from schematic-driven layout methodology that manually determines a waveguide routing plan with crossings pre-inserted in the schematic, we \textbf{assume crossings are unknown during placement and only inserted during routing instead}.
To enable a routing-informed global placement process, we will introduce our key innovations in the following section.

\subsection{PIC-Specific Placement Objective Functions}
\label{sec:Objective}

\subsubsection{Asymmetric Bending-Aware Wirelength.} 
Weighted-average wirelength (WA-WL) is widely used in VLSI placement for wirelength cost, which is used to approximate the half-perimeter wirelength (HPWL). 
However, unlike VLSI routing, PICs only have two-pin nets to describe waveguides and use ports instead of pins to connect the waveguides. 
These ports are essentially open-ended waveguide segments that can only be accessed in a specific orientation.
Therefore, HPWL is not an appropriate approximation for the PIC waveguide routes. 
The original WA-WL is symmetric along the x and y-axis. 
However, as shown in Fig.~\ref{fig:angle}, PIC does not prefer sharp turns. 
Instead, we need to use a smooth, curvy waveguide for bending, which is area-consuming and thus harms routability.

\begin{figure}
    \centering
    \includegraphics[width=0.6\columnwidth]{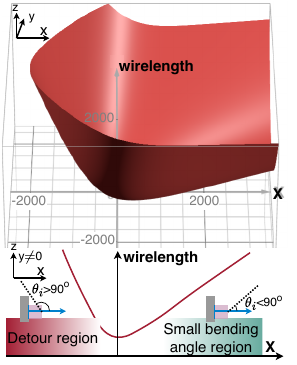}
    \vspace{-10pt}
    \caption{Proposed bending-aware wirelength function: detour region results in larger wirelength cost.}
    \label{fig:wl}
     \vspace{-5pt}
\end{figure}

To be aware of the \emph{port orientation} introduced by inappropriate component locations, and \emph{mitigate the routing resources consumption by curvy bending} in PICs, we propose a \textbf{asymmetric bending-aware wirelength function}, named as $\textrm{cosWA}$ function, as follows,
\begin{equation}
\label{eq:cosWA}
    \begin{aligned}
\mathrm{cosWA}_e &= (1 + W_\theta) \cdot \left( 
\frac{\sum_{i \in e} x_i e^{\frac{x_i}{\gamma}}}{\sum_{i \in e} e^{\frac{x_i}{\gamma}}}
-
\frac{\sum_{i \in e} x_i e^{-\frac{x_i}{\gamma}}}{\sum_{i \in e} e^{-\frac{x_i}{\gamma}}}
\right)^\alpha \\
W_\theta &= [(c-\cos\theta_1)_+]^2 + [(c-\cos\theta_2)_+]^2  \\
\cos\theta_i &= \frac{\mathbf{w} \cdot \mathbf{v_i}}{\|\mathbf{w}\|\|\mathbf{v_i}\|}, \quad i = 1, 2  \\
\mathbf{w} &= (dx, dy), \quad \mathbf{v} \in \{(1, 0), (0, 1), (-1, 0), (0, -1)\} 
\end{aligned}
\end{equation}

\noindent 
Eq.~\eqref{eq:cosWA} shows the horizontal wirelength of net $e$. 
$\vec{w}$ is defined as the vector from the first pin to the second pin of the net, where $dx$ and $dy$ represent the horizontal and vertical distances between the two pins, respectively. 
$\vec{v}$ is the unit vector indicating the port orientation.
Then $\theta_i$ represents the angle between the port orientation and the straight line connection of two-pin nets.
In this way, a large $\theta$ (e.g., 150-degree) will result in a large 
penalty factor $W_\theta$, so that we can penalize the net with a large bending angle as shown in Fig.~\ref{fig:wl}. 
$(\cdot)_+$ is the ReLU function that casts a negative value to 0, which makes sure no extra penalty is added for acute angles. The parameter $c\in [0,1]$ is used to control the angle margin. 
The $W_\theta$ will become 0 when both $\theta_1$ and $\theta_2$ are within the angle margin defined by $c$.
The $\mathrm{cosWA}$ function is asymmetrical along the x-axis if the corresponding port has a horizontal orientation.
Note that the y-axis \textrm{cosWA} wirelength in this case will degrade to a symmetric WA wirelength.
Similarly, it simply flips to x-symmetry and y-asymmetry for vertical ports.

Another key variable here is the exponent $\alpha$ that makes this function not an approximation of linear HPWL, which leads to very long wires and is unable to encourage symmetric layouts, as many placement solutions can have the same HPWL.
To borrow the \emph{long-wire penalty property} of quadratic wirelength without overestimating long waveguides and ignoring short waveguides, we add this $\alpha\in[1,2]$ exponent to make it smoother.
Empirically, we set $\alpha$ to 1.4, which is later identified as the optimal setting in the ablation study section.

\begin{figure}
    \centering
    \includegraphics[width=0.9\columnwidth]{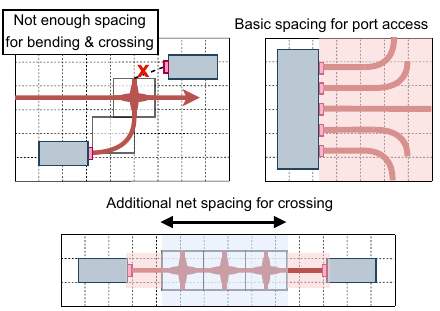}
    \vspace{-10pt}
    \caption{Illustration of the critical spacing requirements in PIC routing that should be considered during placement.}
    \label{fig:SpacingModel}
     \vspace{-10pt}
\end{figure}

\subsubsection{Routing-Informed Net Spacing Model.}
A major root cause of routing failure in PICs is \textbf{spacing deficiency}.
Unlike VLSI, where multiple metal layers are available, PICs are typically limited to a single optical routing layer. 
As a result, routing resources are scarce, and waveguide crossings and curved bends consume significant area.
Without enough spacing, routing is likely to fail (see Fig.~\ref{fig:SpacingModel}(a)).
Hence, it is essential to integrate \emph{routing-informed spacing estimation} into placement. 
Accurate modeling of the spatial demands from port density, bend radii, and anticipated crossings can preempt routing congestion and ensure sufficient space for successful waveguide routing.

Informed by empirical analysis of the routing behavior for the open-source PIC router \texttt{LiDAR}, the \textbf{main routing failure comes from the port accessibility and crossing insertion}, we propose the routing-informed net spacing model $S_i$ based on the port density $P_{\text{dens}}$ and routing congestion $R_{\text{cong}}$.
For each net, we define a \emph{spacing demand} based on two factors: (1) the \textbf{local port density}, which captures how many co-directional ports are clustered, and (2) \textbf{estimated routing congestion}, which accounts for potential waveguide crossings.
Specially, for a 2-pin net $i$ connecting port $p_1$ (on node $v_1$) and port $p_2$ (on node $v_2$), the net spacing demand $S_i$ is estimated as
\begin{equation}
    \label{eq:SpacingModel}
    \begin{aligned}
    S_i&=\max(S_i(v_1, p_1), S_i(v_2, p_2)),\\
        S_i(v, p) &= P_{\text{dens}}(v, p) + R_{\text{cong}}\\
        P_{\text{dens}}(v, p) &= r_{\text{bend}} + \frac{1}{2}P_{\text{num}}(v,p) \times S_{\text{crs}},\\
        P_{\text{num}}(v,p)&=\big|\{p_j\in\text{Ports of }v~~|~~\text{Dir}(p_j)=\text{Dir}(p)\}\big|.
    \end{aligned}
\end{equation}
where the local port density $P_{\text{dens}}$ represents the basic spacing needed for feasible port access, as shown in Fig.~\ref{fig:SpacingModel}(b).
The $r_{\text{bend}}$ is the bending radius, e.g., 5 $\mu m$. 
$P_{\text{num}}(v,p)$ is the number of ports on node $v$ that have the same orientation as port $p$ connected to net $i$.
$S_{\text{crs}}$ is the waveguide crossing height/width (typically a 4-port square-shaped device), e.g., 10 $\mu m$. 
Since a net connects two ports of different components, which gives different port density spacing for the net, we choose the one with the largest required spacing.

Apart from the basic spacing $P_{\text{dens}}$, net crossings also cost a large amount of routing resources that further cause congestion, which should be considered in the spacing model.
Once the positions of all components are roughly stabilized, we estimate the potential crossings to calculate the additional spacing required. 
We estimate the crossings by calculating the line segment intersections between nets during placement. 
If a crossing occurs, we approximate the additional spacing needed for cascaded crossings (see Fig.~\ref{fig:SpacingModel}(c)) by multiplying the number of estimated crossings $\#CR_{net}$ by the size of a crossing 
$S_{\text{crs}} $,
$$R_{\text{cong}} = \#CR_{net} \times S_{\text{crs}} $$
Since net crossings $\#CR_{net}$ evolve as components move, the congestion estimate is updated periodically every 100 iterations starting from the 100\textsuperscript{th} iteration to improve accuracy.
Finally, we define the net spacing penalty for a net $e$ as,
\begin{equation}
    \label{eq:SpacingPenalty}
    NS_e(x,y) = [(v_x\cdot dx-S_i)_+]^2 + [(v_y\cdot dy-S_i)_+]^2,
\end{equation}
where $(v_x, v_y)$ is the port's direction vector, which determines the port's orientation.
Then we calculate the spacing needed for the x-axis and y-axis using a quadratic function. 
$(\cdot)_+$ is $\text{max}(\cdot,0)$ that makes sure we only penalize when spacing is below the threshold.
This penalty term effectively encourages the placer to reserve sufficient space for waveguide routing throughout the layout process.

\subsubsection{Modified Augmented Lagrangian for Electrostatic Density.}
We follow the density model of DREAMPlace3.0~\cite{PD_ICCAD2020_Gu} to remove the overlap among components defined as $D(x,y)$, which models components as electric charges, density as potential energy, and density gradient as electric field. 
The electric potential and field distribution are obtained by solving Poisson’s equation from the charge density distribution via spectral methods with a two-dimensional fast Fourier transform (FFT).
To accelerate the cell spreading and avoid complete cell overlapping that can hardly be separated by this  discretized electrical field, we employ a modified augmented Lagrangian method with a quadratic density term modulated by the density weight $\lambda_D$ to \emph{avoid overly fast spreading in the early stages} that is critical to wirelength optimization,
\begin{equation}
    \label{eq:Density}
    \mathcal{D}(x,y)=\lambda_D\cdot (D(x,y)+\frac{1}{2}\rho D^2(x,y)),
\end{equation}
where $\rho$ is the quadratic penalty coefficient, empirically set to 2,000.
The scheduling of $\lambda_D$ follows the overflow-based schedule~\cite{PLACE_TCAD2020_Lin}.

\noindent\textbf{Filler Size Setting}.~
Fillers are dummy cells introduced to occupy whitespace and maintain electrical equilibrium.
The size of fillers directly influences the granularity of the placement grid and thus affects overall placement precision.
In DREAMPlace, filler sizes are estimated as the average standard cell dimensions. 
However, this assumption breaks down in mixed-size PIC designs, where component geometries vary significantly. 
To better suit the photonic context, we compute the filler aspect ratio based on the average aspect ratio of movable photonic components and clamp it within a predefined range (e.g., [0.2, 5.0]) to avoid extreme shapes. 
Additionally, we flip the filler orientation \emph{opposite to the predominant optical signal flow} to further improve placement adaptability.
In our benchmarks, optical signals primarily propagate from left to right, making placement accuracy in the $x$ direction especially critical. Therefore, we adopt \emph{tall and narrow fillers} to increase resolution along the $x$-axis, enabling fine-grained control over component placement.

\subsubsection{Overall Placement Objective Function.}
In \name, the overall combined objective function is as follows,
\begin{equation}
    \label{eq:Objective}
    \mathcal{L}(x, y) = \sum_{e\in E}\textrm{cosWA}_e(x, y) + \lambda_{NS}\sum_{e\in E}NS_e(x,y) + \mathcal{D}(x, y),
\end{equation}
where $E$ denotes the set of all nets, and $\lambda_{NS}$ is the weight of the spacing penalty, which is set to 1 in the later experiment.
Initially, all components are randomly placed at the layout center, overlapping each other. 
Guided by wirelength minimization and density control, components progressively move toward optimized positions, while the spacing penalty pushes apart those in congested areas to meet spacing constraints.

\subsection{Designer-Constrained Placement via Conditional Projection and Cell Inflation}

Design constraints provided by photonic circuit designers often capture domain-specific knowledge that is crucial for ensuring functional and robust layouts.
For example, some thermo-optic phase shifters should have enough spacing around them to avoid circuit performance degradation caused by the thermal crosstalk. 
Some components or nets should have an almost identical routing topology to achieve phase matching.
However, manually enforcing these constraints is tedious and error-prone, especially under tight geometric constraints.
Our placement engine supports these constraints automatically using two complementary mechanisms: (1) cell inflation for spacing-driven constraints, and (2) projected gradient descent for alignment and uniformity constraints.

\noindent\textbf{Case Study 1: Alignment and Regularity via Conditional Projected Gradient Descent}.~
Many coherent photonic circuits require phase-matched routing or geometric symmetry for robustness considerations.
A common practice is to align such components in rows or grids, ensuring identical optical path lengths and reducing bending complexity. 
Additionally, uniformly distributed components help reserve adequate spacing between waveguides to minimize unwanted coupling and crosstalk. 
Moreover, a regular, evenly spaced layout is inherently more robust to fabrication variations, boosting overall yield and making it far easier to scale designs to larger arrays without sacrificing performance.

\begin{figure}
    \centering
    \includegraphics[width=0.65\columnwidth]{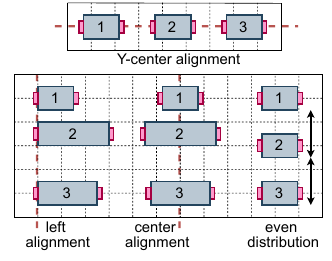}
    \vspace{-10pt}
    \caption{Common physical design constraints in PICs.}
    \label{fig:constraints}
     \vspace{-10pt}
\end{figure}

We enforce such constraints using a conditional projected gradient descent scheme. 
In our PIC benchmark suite, we extend the LEF/DEF syntax to support group-based constraints, e.g., \{\texttt{alignment}: '\texttt{left}', $n_1$, $n_2$, $n_5$\} describes left edge alignment among a group of three nodes.
Figure~\ref{fig:constraints} illustrates different constraints, i.e., left, X-center, and Y-center alignment and uniform spacing.

During placement, nodes in each constraint group are progressively projected toward their feasible locations to honor the constraints. 
To avoid overly hard projection to the feasible set that hinders optimization, we relax the constraints and progressively enforce them using conditional projected gradient descent (CPGD).
We handle these constraints using a projection-based scheme integrated into the gradient descent optimization. Specifically, we adopt a \emph{projection scheduler} that controls the \emph{projection sharpness} of a set of compound projection operators. 
At each iteration $t$ (out of a total $T$ steps), the updated position of each component $x_i$ is computed as a smooth interpolation between its current location and the projected target $x_{\text{new}, i}$:
\begin{equation}
    x_i = (1 - s_t) \cdot x_i + s_t \cdot x_{\text{new}, i},
\end{equation}
where $s_t \in [0,1]$ is an iteration-dependent sharpness value that increases over the course of optimization. 
Here, we use a cosine function to gradually increase projection strength:
\begin{equation}
    s_t = s_0 + (s_T - s_0) \cdot \frac{1 - \cos\left( \pi t / T \right)}{2},
\end{equation}
where $s_0$ and $s_T$ denote the initial and final sharpness, respectively.
For each constraint group $G$ with component positions $\{x_i\}_{i \in G}$, the projected target position $x_{\text{new}, i}$ is computed based on the type of constraint. 
For \emph{alignment constraints}, all components are projected to their group centroid:
\begin{equation}
    x_{\text{new}, i} = \frac{1}{|G|} \sum_{j \in G} x_j.
\end{equation}

For \emph{quantization (uniform spacing)} constraints, each component is snapped to a regular grid within the group's bounding box range:
\begin{equation}
    x_{\text{new}, i} = x_{\min} + s_x \cdot \text{round} \left( \frac{x_i - x_{\min}}{s_x} \right), \quad s_x = \frac{x_{\max} - x_{\min}}{|G| - 1}.
\end{equation}
As optimization progresses, the increasing sharpness $s_t$ ensures that constraint satisfaction is gradually enforced, allowing initial placement flexibility while converging smoothly to the final aligned or evenly spaced configuration.
Note that this projection is not performed after the optimizer descent step. 
Otherwise, it is not aware of this projection.
The projection is applied after the tentative descent for step size estimation in the Nesterov optimizer (see Alg.~\ref{alg:Optimizer} Line 18).

\noindent\textbf{Case Study 2: Spacing Constraint via Cell Inflation}.~
A common constraint is to protect thermo-optic phase shifters from causing thermal crosstalk by enforcing minimum spacings from other cells, e.g., 50 $\mu m$.
This can be converted to an overlap removal problem and elegantly solved by density minimization once the protected cell is inflated with a $50 \mu m$ \emph{halo} (as defined in the LEF/DEF syntax).

\subsection{Stabilizing Mixed-Size PIC Placement via Blockwise Adaptive Nesterov Optimizer}
As emphasized in Section~\ref{sec:Challenge}, the significant heterogeneity in component sizes presents a major challenge for stability and convergence in PIC placement. 
To address this, we introduce a \emph{Blockwise-Adaptive Barzilai–Borwein (BBB)} step size scheme in Nesterov-accelerated gradient descent optimizer in Alg.~\ref{alg:Optimizer}, named BNAG, specifically tailored for mixed-size PIC layouts.
In our formulation, all placement variables are partitioned into 4 logical blocks: movable instances and dummy fillers in both the $x$ and $y$ dimensions. 
An independent Barzilai–Borwein~\cite{PD_ICCAD2023_Chen} step size is computed for each block, and the resulting values are truncated to prevent instability. This blockwise treatment allows us to stabilize the motion of large nodes by decoupling their updates from smaller cells and encourages faster convergence of filler cells to surround movable instances and fill their spacing.
While the Barzilai–Borwein (or inverse Lipschitz) step size is effective for acceleration, it often overestimates the appropriate step size, especially when certain cells are \emph{already near-optimal}. 
To mitigate this and ensure convergence, we apply a global \emph{cosine annealing schedule} to gradually reduce the effective step size over time.
This helps maintain robustness in the presence of highly uneven cell dimensions and density.
Additionally, we project $v_{k+1}$ onto the feasible placement region to correct for potential extrapolation beyond legal bounds in Nesterov’s momentum. This guarantees that gradient evaluations occur at valid locations, preserving placement legality and improving stability.

\begin{algorithm}
\caption{Blockwise Adaptive Nesterov-accelerated Gradient Descent (BNAG) Optimizer (one step).}
\label{alg:Optimizer}
\begin{algorithmic}[1]
\Require
    $a_k$ (optimization parameter), $u_k$ (major solution), $v_k$ (reference solution), $v_{k-1}$, $\nabla f(v_k)$, $\nabla f(v_{k-1})$, $\{B_1,\dots,B_m\}$
\Ensure 
  $u_{k+1} (\text{updated sol.}),\;v_{k+1},\;a_{k+1}$
\For{$j = 1$ \textbf{to} $m$}
  \State $g^{(k-1)} \gets \nabla f(v_{k-1})[B_j]$
  \State $g^{(k)} \gets \nabla f(v_k)[B_j]$
  \State $s^{(k-1)} \gets v_k[B_j] - v_{k-1}[B_j]$
  \State $y^{(k-1)} \gets g^{(k)} - g^{(k-1)}$
  \State $ \alpha_{\text{bb}}^{(j)} \gets \frac{(s^{(k-1)})^T y^{(k-1)}}{(y^{(k-1)})^T y^{(k-1)}} $
  \State $ \alpha_{\text{lip}}^{(j)} \gets \frac{\|s^{(k-1)}\|}{\|y^{(k-1)}\|} $
  \State \textbf{if} $ \alpha_{\text{bb}}^{(j)} > 0 $ \textbf{then} $\alpha_k^{(j)} \gets \alpha_{\text{bb}}^{(j)}$
  \State \textbf{else} $\alpha_k^{(j)} \gets \min(\alpha_{\text{lip}}^{(j)},\; \alpha_{k-1}^{(j)})$
  \State $\alpha_k^{(j)} \gets \alpha_k^{(j)} \cdot \eta_k$
\EndFor
\For{each block $j = 1, \dots, m$ \textbf{and} each $i \in B_j$}
  \State $u_{k+1}[i] \gets v_k[i] - \alpha_k^{(j)} \cdot \nabla f(v_k)[i]$
\EndFor
\State $a_{k+1} \gets \frac{1 + \sqrt{4a_k^2 + 1}}{2}$
\State $v_{k+1} \gets u_{k+1} + \frac{a_k - 1}{a_{k+1}} (u_{k+1} - u_k)$
\State $\eta_{k+1} \gets \eta_{\min} + \frac{1}{2} (\eta_0 - \eta_{\min}) \left(1 + \cos\left(\frac{\pi k}{K_{\max}}\right)\right)$
\State \textbf{project}$(v_{k+1})\quad\triangleright$ enforce constraints
\State \Return $u_{k+1},\;v_{k+1},\;a_{k+1}$
\end{algorithmic}
\end{algorithm}

\subsection{Legalization}
\label{legallization}
Since the components in the PICs have a relatively large feature size, we directly perform macro legalization~\cite{PLACE_TCAD2020_Lin} after the global placement. 
The greedy Macro Legalization operator legalizes movable macros in two stages: first, a coarse Hanan-grid pass for small clusters and blocked macros, and then an LP/graph-based refinement, alternating until total and weighted displacement are minimized and the best legal layout is retained. 

\begin{figure}
    \centering
    \includegraphics[width=\columnwidth]{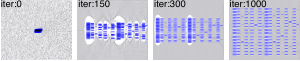}
    \vspace{-15pt}
    \caption{Placement animation of \name on ADEPT\_16x16.}
    \label{fig:animation}
     \vspace{-10pt}
\end{figure}

We show an \textbf{animation of the placement process} of \name on a representative benchmark to visualize its behavior in Fig.~\ref{fig:animation}.

\section{Evaluation Results}
\label{sec:ExperimentalResults}
The \name framework builds upon DREAMPlace, with PIC-specific optimization.  
All the experiments are conducted on an AMD EPYC 7763 Linux server with a 2.9GHz CPU and NVIDIA RTX A6000 GPU.

\begin{figure}
    \centering
    \includegraphics[width=1\columnwidth]{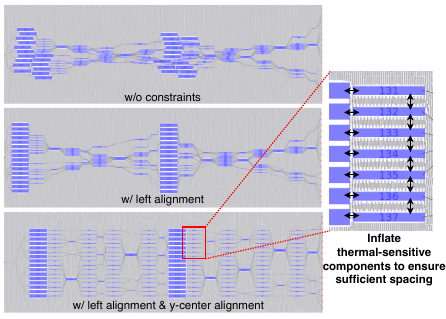}
    \vspace{-15pt}
    \caption{Proper placement constraints ensure high PIC layout quality on ADEPT\_16$\times$16 circuit.}
    \label{fig:placement}
     \vspace{-10pt}
\end{figure}

\noindent\textbf{Benchmarks}.~In our customized realistic benchmark suites, we use the Clements-style Mach-Zehnder Interferometer (MZI)~\cite{NP_NATURE2017_Shen} arrays and an auto-searched photonic tensor core design (ADEPT)~\cite{NP_DAC2022_Gu} at two different scales. 
For each circuit, we generate two chip configurations:
$S$ size with a very compact die size with a 5 $\mu m$ bending radius and 
$L$ size with a more relaxed die size with a 10 $\mu m$ bending radius, and use a 10$\mu m$ $\times$10$\mu m$ crossing size.

\noindent\textbf{Baselines}.~We compare \name against two baselines:
DREAMPlace+RO, DREAMPlace augmented with routability optimization via node inflation based on the RUDY congestion map, and a PCB placement tool named Cypress~\cite{PD_ISPD2025_Zhang} that minimizes net crossings to improve routability. 

\noindent\textbf{Evaluation Metrics}.~We evaluate the placement solution quality using several key metrics: \textbf{routability,  number of crossings (\#CR), and runtime}. 
Routability is defined as the ratio of successfully routed nets to the total number of nets. 
To evaluate routability, we perform exact waveguide routing on the legalized placement using an open‑source PIC router LiDAR~\cite{PD_ISPD2025_Zhou}.
Crossing counts, as an indirect routability indicator, are computed directly from the placement by detecting intersections of the straight‑line segments representing each two‑pin net.
Insertion loss $IL_{max}$ is a critical metric related to waveguide bending angles, routing length, and crossings that implies the optical signal integrity.
Note that without a fully DRV-free layout, it is not meaningful to evaluate post-routing insertion loss.
Instead, we estimate the \textbf{pre-routing insertion loss} $IL_{max}$ after the placement stage from the estimated routing length and port orientations, predicting bending angles and applying an analytical waveguide insertion loss model~\cite{PD_ISPD2025_Zhou}.

\subsection{Handling of Design Constraints}

Constraints are absolutely critical in PIC placement. Due to the complex port alignments in PIC circuits, unreasonable device placement can severely degrade routability, make it impossible to produce a legal layout, induce thermal crosstalk, and exacerbate manufacturing variability. Therefore, it is essential to impose layout constraints from the very beginning. Figure~\ref{fig:placement} illustrates how placement evolves as left‑alignment and y‑center‑alignment constraints are gradually applied. With no constraints, components are placed in a highly irregular fashion, leading to severe port misalignment and an entirely unrouteable solution. As constraints accumulate, the circuit becomes increasingly regular: grouping certain components for left alignment arranges them into neat columns, though the groups remain too close and can introduce crosstalk; adding y‑center alignment then transforms the layout into a true grid, perfectly aligning ports and markedly enhancing routability. To ensure a fair comparison in our experiments, we apply the same constraint set to all baseline methods so that each can produce a valid, reasonable placement.

\subsection{Main Results}
\begin{table*}[t]
\caption{Comparison of the routability, the estimated number of crossings after placement, and runtime (s).}
\vspace{-8pt}
\resizebox{0.9\textwidth}{!}{
\begin{tabular}{|cccccccccc|}
\hline
\multirow{2}{*}{Benchmark} & \multicolumn{3}{c}{DREAMPlace /w   routability optimization~\cite{PLACE_TCAD2020_Lin}} & \multicolumn{3}{c}{Cypress~\cite{PD_ISPD2025_Zhang}}      & \multicolumn{3}{c|}{\name}       \\ \cline{2-10} 
                           & \#CR           & Routability          & Runtime (s)          & \#CR & Routability & Runtime (s) & \#CR & Routability & Runtime (s) \\ \hline
Clements\_8$\times$8\_S           & 7              & 43.68\%              & 30.15                & 0    & 72.41\%     & 15.80       & 0    & 98.85\%     & 21.13       \\
Clements\_8$\times$8\_L           & 8              & 50.42\%              & 31.64                & 0    & 77.73\%     & 16.48       & 0    & 98.85\%     & 23.07       \\
Clements\_16$\times$16\_S         & 20             & 63.70\%              & 43.67                & 13   & 41.91\%     & 22.33       & 2    & 95.05\%     & 31.89       \\
Clements\_16$\times$16\_L         & 28             & 59.43\%              & 45.19                & 17   & 48.64\%     & 23.01       & 2    & 96.04\%     & 32.64       \\
ADEPT\_8$\times$8\_S              & 38             & 48.82\%              & 39.64                & 61   & 42.73\%     & 21.58       & 28   & 90.05\%     & 22.79       \\
ADEPT\_8$\times$8\_L              & 49             & 70.63\%              & 40.92                & 53   & 61.85\%     & 22.63       & 33   & 95.04\%     & 23.11       \\
ADEPT\_16$\times$16\_S            & 83             & 47.96\%              & 61.57                & 105  & 35.42\%     & 34.36       & 79   & 98.12\%     & 35.88       \\
ADEPT\_16$\times$16\_L            & 124            & 40.87\%              & 63.24                & 113  & 33.79\%     & 35.62       & 87   & 90.60\%     & 39.62       \\
ADEPT\_32$\times$32\_S            & 261            & 44.20\%              & 117.41               & 209  & 62.71\%     & 64.43       & 206  & 96.74\%     & 65.99       \\
ADEPT\_32$\times$32\_L            & 263            & 48.37\%              & 124.73               & 253  & 46.74\%     & 69.89       & 237  & 93.38\%     & 71.62       \\
ADEPT\_64$\times$64\_S            & 381            & 43.89\%              & 235.31               & 358  & 41.82\%     & 133.39      & 314  & 97.10\%     & 105.96      \\
ADEPT\_64$\times$64\_L            & 303            & 48.17\%              & 238.47               & 297  & 50.83\%     & 136.71      & 209  & 87.62\%     & 110.93      \\ \hline
Geo-mean                   & -              & 50.85\%              & 89.33                & -    & 51.38\%     & 49.69       & -    & 94.79\%     & 48.72       \\
Ratio                      & -              & 0.53                 & 1.83                 & -    & 0.54        & 1.02        & -    & 1           & 1           \\ \hline
\end{tabular}
}
\label{tab:main_result}
\vspace{-5pt}
\end{table*}

\begin{figure}
    \centering
    \includegraphics[width=0.75\columnwidth]{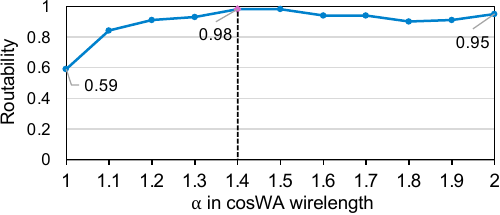}
    \vspace{-10pt}
    \caption{Impact of the exponent $\alpha$ in our cosWA wirelength on routability on ADEPT\_16$\times$16.}
    \label{fig:alpha}
     \vspace{-10pt}
\end{figure}

\begin{table}[]
\caption{Random initialization vs. manual initial placement. 
\name is robust and does not rely on good initial placement.
}
\resizebox{0.9\columnwidth}{!}{
\begin{tabular}{|c|cc|cc|}
\hline
\multirow{2}{*}{Benchmarks} & \multicolumn{2}{c|}{Random initialization} & \multicolumn{2}{c|}{Manual initialization} \\ \cline{2-5} 
                            & \#CR             & Routability $\uparrow$             & \#CR             & Routability $\uparrow$             \\ \hline
Clements\_16$\times$16\_S             & 2                & 95.05\%                 & 2                & 95.05\%                 \\ \hline
ADEPT\_32$\times$32\_S                & 206              & 96.74\%                 & 183              & 95.36\%                 \\ \hline
ADEPT\_64$\times$64\_S                & 314              & 97.10\%                 & 304              & 96.32\%                 \\ \hline
\end{tabular}
}
\label{tab:init}
\vspace{-10pt}
\end{table}

\begin{figure*}
    \centering
    \includegraphics[width=0.82\textwidth]{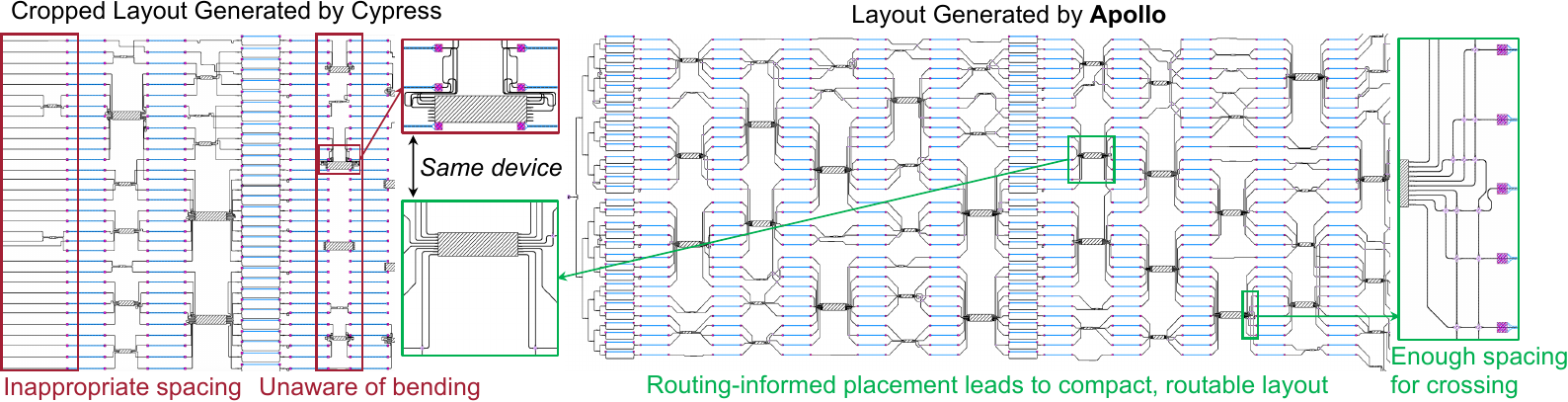}
    \vspace{-8pt}
    \caption{ADEPT\_32$\times$32 visual comparison of post-routing layout (viewed in KLayout) generated by Cypress and our \name.}
    \label{fig:kayout_vis}
     \vspace{-8pt}
\end{figure*}

For the parameters in the proposed cosWA function, we first performed a parameter sweep on the ADEPT\_16x16 benchmark. As shown in Fig.~\ref{fig:alpha}, we varied $\alpha$ from 1.0 to 2.0 in increments of 0.1, ran placement followed by full routing at each setting, and recorded the resulting routability. We then selected $\alpha=1.4$, the value that produced the highest routability for all subsequent experiments. As for the margin parameter $c$ in cosWA, because we employ Manhattan routing in later stages, we set c=0; that is, once the ports achieve a 90-degree orientation, no further bending‐angle penalty is applied.

Moreover, in all subsequent experiments, we initialize every component at a random center location. As Table~\ref{tab:init} shows, even when compared to manually placing components in favorable starting positions, our method still converges to high‑quality solutions under random center initialization, demonstrating the robustness of the proposed approach.

Table~\ref{tab:main_result} shows that \name consistently outperforms two baselines, four Clements and four ADEPT benchmarks, under both compact and relaxed die sizes. 
\name achieves a geometric mean routability of 94.79\%, versus 50.85\% for DREAMPlace and 51.38\% for Cypress. 
In all S‑size runs, \name exceeds 98.5\% routability, drives net crossings down to zero on Clements\_8$\times$8\_S and Clements\_16$\times$16\_S, and finishes in under 25s on average, roughly 1.8$\times$ faster than DREAMPlace and on par with Cypress. 
Even on the largest ADEPT\_64$\times$64\_L circuit, \name achieves 87.6\% routability, significantly outperforming prior methods ($\sim$48\%/51\%), and reduces waveguide crossings by $\sim$30\%. 
Despite the L benchmark offering a larger area, its 10 $\mu m$ bending radius makes routing more challenging than in the S benchmark. 
This illustrates that a larger die size does not suffice to resolve local congestion issues.

Unlike the baselines, which optimize a generic VLSI wirelength objective, \name is designed specifically for PIC placement and explicitly accounts for the spatial footprint of curved waveguide bends and crossings. 
This routing‑aware formulation yields post‑placement layouts that are inherently highly routable. 
Even though DREAMPlace+RO performs node inflation to create extra port spacing based on congestion maps, it does not capture the area overhead of curvy bends, leaving many nets difficult to access and route.

\subsection{Ablation Studies}
\subsubsection{Wirelength Function Comparison}
\begin{table*}[]
\caption{Comparison of total bending angle $BA_{tot}$ and max insertion loss $IL_{max}$ for different wirelength models.
($\downarrow$): lower is better.}
\vspace{-5pt}
\resizebox{0.75\textwidth}{!}{
\begin{tabular}{|c|cc|cc|cc|cc|}
\hline
\multirow{2}{*}{Benchmark} & \multicolumn{2}{c|}{WA-WL}   & \multicolumn{2}{c|}{LSE}     & \multicolumn{2}{c|}{Quadratic} & \multicolumn{2}{c|}{\textbf{Proposed cosWA}}        \\ \cline{2-9} 
                           & $BA_{tot}\downarrow$ & $IL_{max}\downarrow$    & $BA_{tot}\downarrow$ & $IL_{max}\downarrow$    & $BA_{tot}\downarrow$    & $IL_{max}\downarrow$   & $BA_{tot}\downarrow$      & $IL_{max}\downarrow$    \\ \hline
Clements\_8$\times$8\_S              & 1.64E+04     & 0.32          & 1.57E+04     & \textbf{0.30} & 1.66E+04        & 0.33         & \textbf{1.57E+04} & 0.31          \\
Clements\_16$\times$16\_S              & 5.61E+04     & \textbf{0.59} & 8.24E+04     & 0.73          & 5.63E+04        & 0.60         & \textbf{5.37E+04} & 0.60          \\
ADEPT\_8$\times$8\_S                 & 2.44E+04     & 0.61          & 3.09E+04     & 0.67          & 2.40E+04        & 0.60         & \textbf{2.06E+04} & \textbf{0.57} \\
ADEPT\_16$\times$16\_S               & 5.96E+04     & 1.31          & 7.49E+04     & 1.37          & 5.62E+04        & 1.28         & \textbf{5.17E+04} & \textbf{1.25} \\
ADEPT\_32$\times$32\_S               & 1.63E+05     & 2.58          & 1.74E+05     & 2.46          & 1.60E+05        & 2.44         & \textbf{1.17E+05} & \textbf{2.32} \\
ADEPT\_64$\times$64\_S               & 2.74E+05     & 3.30          & 3.17E+05     & 3.43          & 2.96E+05        & 3.04         & \textbf{2.44E+05} & \textbf{2.89} \\ \hline
Geo-mean                   & 9.90E+04     & 1.45          & 1.16E+05     & 1.49          & 1.01E+05        & 1.38         & \textbf{8.37E+04} & \textbf{1.32} \\
Ratio                      & 1.18         & 1.10          & 1.38         & 1.23          & 1.19            & 1.05         & \textbf{1.00}     & \textbf{1.00} \\ \hline
\end{tabular}
}
\label{tab:wl_objective_comparison}
\vspace{-8pt}
\end{table*}

To evaluate the impact of different wirelength models on photonic layout quality, we compare weighted‑average wirelength (WA‑WL)~\cite{WA-WL}, log‑sum‑exponential (LSE)~\cite{LSE-WL}, quadratic wirelength, and our proposed cosine‑weighted wirelength (cosWA). 
As shown in Table~\ref{tab:wl_objective_comparison}, cosWA achieves the lowest maximum insertion loss ($IL_{\max}$), outperforming WA‑WL, LSE, and quadratic by 1.10$\times$, 1.13$\times$, and 1.05$\times$, respectively. 
It also produces the smallest total bending angle improvements of roughly 1.18$\times$, 1.38$\times$, and 1.21$\times$ over WA‑WL, LSE, and quadratic wirelength.
These reductions in loss and detour translate directly into \emph{compact layouts, reduced local congestion, and ultimately significantly improved routability}.

\subsubsection{Spacing Model Comparison}
\begin{table*}[!t]
\centering
\caption{Routing success rates (higher the better) under different placement spacing models (w/o spacing, port-density-based component inflation ($PI$) and our spacing model with different configuration) on various benchmarks (S is small chip space and L is large chip space).
Our net spacing model, based on joint waveguide, bending, and crossing modeling (last column), leads to the best routing success rate.
\emph{PI}: cell inflation based on pin density.
}
\vspace{-8pt}
\scriptsize                      %
\setlength\tabcolsep{3pt}       %
\resizebox{1.3\columnwidth}{!}{%
\begin{tabular}{|cccccc|}
\hline
\multicolumn{1}{|c|}{\multirow{2}{*}{Benchmark}} & \multicolumn{1}{c|}{\multirow{2}{*}{w/o spacing}} & \multicolumn{1}{c|}{\multirow{2}{*}{$PI$}} & \multicolumn{2}{c|}{Only basic spacing ($P_{\text{dens}}$)}                            & \multirow{2}{*}{\textbf{$P_{\text{dens}}+R_{\text{cong}}$}} \\ \cline{4-5}
\multicolumn{1}{|c|}{}                           & \multicolumn{1}{c|}{}                                   & \multicolumn{1}{c|}{}                                                   & \multicolumn{1}{c|}{$r_{\text{bend}}$} & \multicolumn{1}{c|} {$P_{\text{num}}\times S_{\text{crs}}$} &                                                               \\ \hline
\multicolumn{1}{|c|}{Clements\_8$\times$8\_S}           & \multicolumn{1}{c|}{94.25\%}                            & \multicolumn{1}{c|}{95.40\%}                                            & \multicolumn{1}{c|}{95.40\%}                    & \multicolumn{1}{c|}{94.25\%}                 & 98.85\%                                                       \\
\multicolumn{1}{|c|}{Clements\_8$\times$8\_L}           & \multicolumn{1}{c|}{93.01\%}                            & \multicolumn{1}{c|}{96.43\%}                                            & \multicolumn{1}{c|}{95.54\%}                    & \multicolumn{1}{c|}{97.70\%}                 & 98.85\%                                                       \\
\multicolumn{1}{|c|}{Clements\_16$\times$16\_S}         & \multicolumn{1}{c|}{82.51\%}                            & \multicolumn{1}{c|}{85.81\%}                                            & \multicolumn{1}{c|}{85.48\%}                    & \multicolumn{1}{c|}{87.13\%}                 & 95.05\%                                                       \\
\multicolumn{1}{|c|}{Clements\_16$\times$16\_L}         & \multicolumn{1}{c|}{74.59\%}                            & \multicolumn{1}{c|}{85.15\%}                                            & \multicolumn{1}{c|}{87.13\%}                    & \multicolumn{1}{c|}{93.07\%}                 & 96.04\%                                                       \\
\multicolumn{1}{|c|}{ADEPT\_8$\times$8\_S}              & \multicolumn{1}{c|}{76.38\%}                            & \multicolumn{1}{c|}{78.74\%}                                            & \multicolumn{1}{c|}{70.87\%}                    & \multicolumn{1}{c|}{87.40\%}                 & 90.05\%                                                       \\
\multicolumn{1}{|c|}{ADEPT\_8$\times$8\_L}              & \multicolumn{1}{c|}{65.80\%}                            & \multicolumn{1}{c|}{70.08\%}                                            & \multicolumn{1}{c|}{76.38\%}                    & \multicolumn{1}{c|}{87.74\%}                 & 95.04\%                                                       \\
\multicolumn{1}{|c|}{ADEPT\_16$\times$16\_S}            & \multicolumn{1}{c|}{60.50\%}                            & \multicolumn{1}{c|}{61.76\%}                                            & \multicolumn{1}{c|}{54.23\%}                    & \multicolumn{1}{c|}{95.61\%}                 & 98.12\%                                                       \\
\multicolumn{1}{|c|}{ADEPT\_16$\times$16\_L}            & \multicolumn{1}{c|}{63.95\%}                            & \multicolumn{1}{c|}{69.91\%}                                            & \multicolumn{1}{c|}{84.01\%}                    & \multicolumn{1}{c|}{86.21\%}                 & 90.60\%                                                       \\
\multicolumn{1}{|c|}{ADEPT\_32$\times$32\_S}            & \multicolumn{1}{c|}{49.15\%}                            & \multicolumn{1}{c|}{51.89\%}                                            & \multicolumn{1}{c|}{53.85\%}                    & \multicolumn{1}{c|}{96.35\%}                 & 96.74\%                                                       \\
\multicolumn{1}{|c|}{ADEPT\_32$\times$32\_L}            & \multicolumn{1}{c|}{62.40\%}                            & \multicolumn{1}{c|}{63.31\%}                                            & \multicolumn{1}{c|}{72.62\%}                    & \multicolumn{1}{c|}{83.96\%}                 & 93.38\%                                                       \\
\multicolumn{1}{|c|}{ADEPT\_64$\times$64\_S}            & \multicolumn{1}{c|}{40.87\%}                            & \multicolumn{1}{c|}{52.65\%}                                            & \multicolumn{1}{c|}{66.44\%}                    & \multicolumn{1}{c|}{91.29\%}                 & 97.10\%                                                       \\
\multicolumn{1}{|c|}{ADEPT\_64$\times$64\_L}            & \multicolumn{1}{c|}{50.98\%}                            & \multicolumn{1}{c|}{66.32\%}                                            & \multicolumn{1}{c|}{70.77\%}                    & \multicolumn{1}{c|}{76.94\%}                 & 87.62\%                                                       \\ \hline
\multicolumn{1}{|c|}{Geo-mean}                                         & 67.87\%                                                 & 73.12\%                                                                 & 76.06\%                                         & 89.80\%                                      & \textbf{94.79\%}                                                       \\
\multicolumn{1}{|c|}{Ratio}                                            & 0.72                                                    & 0.77                                                                    & 0.80                                             & 0.95                                         & \textbf{1.00}                                                             \\ \hline
\end{tabular}
}
\vspace{-8pt}
\label{tab:spacing-comparison}
\end{table*}

\begin{table*}[]
\caption{Compare our proposed BNAG optimizer to other optimizers used in DREAMPlace. NA means the diverged placement has very low quality and cannot be evaluated for total bending angle ($BA_{tot}$) and maximum insertion loss ($IL_{max}$).
BNAG achieves the most stable convergence and the highest placement quality.
($\downarrow$) means lower is better.
}
\vspace{-8pt}
\resizebox{\textwidth}{!}{
\begin{tabular}{|c|ccc|ccc|ccc|ccc|}
\hline
\multirow{2}{*}{Benchmark} & \multicolumn{3}{c|}{\textbf{Proposed BNAG}}          & \multicolumn{3}{c|}{NAG+BB~\cite{PD_ICCAD2023_Chen}}            & \multicolumn{3}{c|}{NAG~\cite{PLACE_TCAD2020_Lin}}          & \multicolumn{3}{c|}{Adam~\cite{NN_ICLR2015_Kingma}}          \\ \cline{2-13} 
                           & Status  & $BA_{tot} \downarrow$       & $IL_{max}\downarrow$    & Status  & $BA_{tot}\downarrow$       & $IL_{max}\downarrow$    & Status  & $BA_{tot}\downarrow$       & $IL_{max}\downarrow$    & Status  & $BA_{tot}\downarrow$       & $IL_{max}\downarrow$    \\ \hline
Clements\_8$\times$8\_S           & Success & 1.56E+04    & 0.294      & Success & 1.59E+04    & 0.357      & Diverge & 1.61E+04    & 0.307      & Diverge & 1.53E+04    & 0.285      \\
Clements\_16$\times$16\_S         & Success & 5.75E+04    & 0.684      & Success & 5.35E+04    & 0.655      & Diverge & 5.54E+04    & 0.666      & Success & 5.96E+04    & 0.690       \\
ADEPT\_8$\times$8\_S              & Success & 1.78E+04    & 0.464      & Success & 1.81E+04    & 0.467      & Success & 1.95E+04    & 0.724      & Diverge & NA          & NA         \\
ADEPT\_16$\times$16\_S            & Success & 4.18E+04    & 1.097      & Success & 4.59E+04    & 1.610       & Success & 4.62E+03    & 1.540       & Success & 4.42E+04    & 1.363      \\
ADEPT\_32$\times$32\_S            & Success & 1.03E+05    & 1.497      & Success & 1.20E+05    & 2.958      & Diverge & 1.15E+05    & 1.932      & Diverge & NA          & NA         \\
ADEPT\_64$\times$64\_S            & Success & 2.48E+05    & 1.901      & Success & 2.52E+05    & 2.619      & Success & 2.69E+05    & 3.779      & Success & 2.93E+05    & 2.706      \\ \hline
Geo-mean    & - &  8.06E+04     &  0.99     & - & 8.43E+04    &   1.44    & - &7.99E+04      &  1.49     & - & -    & -      \\ 
Ratio      &    -     & 1.00 & 1.00     & -       &1.05  &  1.46    & - &  0.99       & 1.51 &  -    & -     & -       \\
\hline
\end{tabular}
}
\vspace{-10pt}
\label{tab:CompareOptimizer}
\end{table*}

Although cosWA wirelength substantially reduces unnecessary waveguide detours, regions traversed by multiple waveguides can still become heavily congested. 
This is especially true for multi‑port components, where servicing or escaping several ports in a tight area entails numerous bends and crossings—scenarios that conventional VLSI node‑inflation cannot adequately resolve. 
To address this, we borrow from automated waveguide routing practices and introduce explicit spacing models to mitigate congestion arising from both bends and crossings. We conduct an ablation study comparing: (1) no spacing mode, (2) port-density inflation ($PI$): adding Halo to cells proportional to the product of number of ports and bending radius for the cell, our proposed basic net spacing schemes, including two variants: (3) bend-radius spacing ($r_{\text{bend}}$) and (4) the port-count-crossing product $S_{\text{crs}}$, and (5) our final net spacing model with basic spacing and congestion cores ($P_{\text{dens}}+R_{\text{cong}}$).

As Table~\ref{tab:spacing-comparison} shows, port‑density-based cell inflation does improve routability over \emph{no spacing model} (geometric‑mean success rate rises from 67.87\% to 73.12\%), but it still fails on ADEPT circuits with heavy multi‑port crossings. 
Introducing a basic bend‑radius clearance ($r_{\rm bend}$) further boosts the geo‑mean to 76.06\%, yet it still overlooks congestion at port crossings. 
When we explicitly account for crossing‑induced spacing, the routability jumps to 89.80\%. 
Finally, our full net spacing model, combining port‑density and crossing-aware spacing, delivers \textbf{an average success rate of 94.79\% across all benchmarks}.

\subsubsection{Optimizer Comparison}

Table~\ref{tab:CompareOptimizer} compares our proposed BNAG optimizer with NAG+BB~\cite{PD_ICCAD2023_Chen}, NAG~\cite{PLACE_TCAD2020_Lin}, and Adam~\cite{NN_ICLR2015_Kingma} over six benchmarks under the same number of iterations. 
While NAG and Adam optimizers diverge on several test cases, both BNAG and NAG+BB converge across all benchmarks. 
Moreover, BNAG achieves an improvement of approximately \(1.04\times\) in total bending cost and \(1.29\times\) in maximum insertion loss.

\section{Conclusion}
\label{sec:Conclusion}
We introduce \name, the first routing-informed, constraint-aware, GPU-accelerated placement framework for large-scale photonic ICs. 
By modeling waveguide constraints, port orientations, and physical spacing requirements as first-class citizens during placement, \name bridges the longstanding gap in photonic physical design automation toolflow. 
Through a synergy of bending-aware wirelength modeling, congestion-driven spacing, progressive constraint projection, and blockwise adaptive optimization, \name achieves highly routable layouts within minutes, even for PICs with thousands of components and a stringent chip area budget and layout constraints.
In addition to the placement engine, we contribute a suite of realistic, large-scale PIC benchmarks derived from real photonic tensor core designs, fostering reproducibility and future research in layout automation. 
We show that \name consistently outperforms prior approaches in routability, layout quality, and runtime. As photonics becomes central to future computing and interconnects, \name offers a powerful foundation for next-generation electronic-photonic design automation.


\end{document}